\documentclass[12pt,psfig]{article}

\newcommand{\dfcoll}{{\left[ \frac{df}{dt} \right] _{\textrm{coll}}}}
\usepackage{graphicx}
\usepackage{epsf}
\usepackage{amssymb}
\usepackage{amsmath}
\title{Low-variance Monte Carlo Solutions of the Boltzmann Transport Equation}
\author{
Nicolas G. Hadjiconstantinou, Gregg A. Radtke and Lowell L. Baker\\
{\it Mechanical Engineering Department}\\
{\it Massachusetts Institute of Technology}\\
{\it Cambridge, MA 02139}}

\date{\today}

\begin{document}

\maketitle

\begin{abstract}
We present and discuss a variance-reduced stochastic particle method for simulating the relaxation-time model of the 
Boltzmann transport equation. The present paper focuses on the dilute gas case, although the method is expected to directly extend to all fields (carriers) for which the relaxation-time approximation is reasonable. The variance reduction, achieved by simulating only the deviation from equilibrium, results in a  significant computational efficiency advantage compared to traditional stochastic particle methods in the limit of small deviation from equilibrium. More specifically, the proposed method can efficiently simulate arbitrarily small deviations from equilibrium  at a computational cost that is independent of the deviation from  equilibrium, which is in sharp contrast to traditional particle methods. 
\end{abstract}

The Boltzmann transport equation 
\begin{equation}
\frac{\partial f}{\partial t} + \mathbf{c} \cdot \frac{\partial f}{\partial \mathbf{r}} + \mathbf{a} \cdot \frac{\partial f}{\partial \mathbf{c}}= \dfcoll
\end{equation}
where $f({\bf r},{\bf c},t)$ is the single-particle distribution function \cite{cerc}, $\dfcoll$ denotes the collision operator, ${\bf a}=(a_x,a_y,a_z)$ is the acceleration due to an external field, ${\bf r}=(x,y,z)$ is the position vector in physical space, ${\bf c}=(c_x,c_y,c_z)$ is the molecular velocity vector, and $t$ is time, is used to describe (under appropriate conditions) transport processes in a wide variety of fields \cite{gang} including dilute gas flow \cite{cerc}, phonon \cite{phonon}, electron  \cite{electron}, neutron \cite{neutron} and photon transport \cite{photon}. Recently, it has received renewed attention in connection to micro- and nano-scale science and technology where transport at lengthscales of the order of, or smaller than, the carrier mean free path is frequently considered (e.g. nanoscale solid-state heat transfer \cite{ball-diff,majumdar})

Numerical solution of the Boltzmann equation remains a formidable task due to the complexity associated with the collision operator and the high dimensionality of the distribution function.   Both these features have contributed to the prevalence of particle solution methods, which are typically able to simulate the collision operator through simple and physically intuitive stochastic processes while employing importance sampling, which reduces computational cost and memory usage \cite{pof2005}. Another contributing factor to the prevalence of particle schemes is their natural treatment of the advection operator, which results in a numerical method that can  easily handle and accurately capture traveling discontinuities in the distribution function \cite{pof2005}. An example of a particle method is the direct simulation Monte Carlo (DSMC) \cite{bird} which has become the prevalent simulation method for dilute gas flow. 

One of the most important disadvantages of particle methods for solving the Boltzmann equation derives from their reliance on statistical averaging for extracting field quantities from particle data. When simulating processes close to equilibrium, thermal noise typically exceeds the available signal. When coupled with the slow convergence of statistical sampling (statistical error decreases with the square root of the number of samples), this often leads to computationally intractable problems \cite{jcp}. For example, to resolve a flow speed of the order of 1 m/s to 1\% statistical uncertainty in a dilute gas, on the order of $5\times 10^8$ independent samples are needed \cite{jcp}. 

In a recent paper, Baker and Hadjiconstantinou have shown \cite{pof2005} that this rather severe limitation can be overcome with a form of variance reduction achieved by simulating only the deviation from equilibrium. By adopting this approach, it is possible to construct Monte Carlo simulation methods that can capture arbitrarily small deviations from equilibrium at a computational cost that is independent of the magnitude of this deviation. This is in sharp contrast to  regular Monte Carlo methods, such as DSMC, whose computational cost for the same signal-to-noise ratio increases {\it sharply} \cite{jcp} as the deviation from equilibrium decreases.

The work in Refs \cite{pof2005,lowell,thomas1,thomas2} focused on the Boltzmann equation for hard spheres and the associated hard-sphere collision operator. The complexity associated with this collision operator, as well as others in related fields, has prompted scientists to search for simplified models; one particularly popular model is the relaxation-time approximation \cite{cerc,gang} 
\begin{equation}
\dfcoll=-\frac{1}{\tau} \left(f - f^{loc}\right)
\end{equation}
where $f^{loc}$ is the {\it local equilibrium} distribution function and $\tau$ is a relaxation time. Despite the approximation involved, this collision model  has enjoyed widespread application in a variety of disciplines concerned with transport processes \cite{cerc,gang,electron,ball-diff,majumdar,xu,sone}.

In response to this widespread use, in the present paper, we present a variance-reduced particle method for simulating the Boltzmann equation under the relaxation-time approximation.  To focus the discussion, we specialize our treatment to the dilute gas case; however, we hope that this exposition can serve as a prototype for development of similar techniques in all fields where the relaxation-time approximation is applicable. Within the rarefied gas dynamics literature, the relaxation-time approximation is known as the BGK model \cite{cerc}.

In the interest of simplicity, in the present paper we assume $\tau\neq\tau({\bf c})$ and that no external forces are present. The first assumption can be easily relaxed, as discussed below. External fields also require relatively straightforward modifications to the algorithm presented below.

As discussed in previous work \cite{pof2005}, a variance-reduced formulation is obtained by simulating only the deviation $f^d \equiv f - f^e$ from an \emph{arbitrary}, but judiciously chosen, underlying equilibrium distribution $f^e$. In other words,  computational particles represent the deviation from equilibrium and, as a result, they may be positive or negative, depending on the sign of the deviation from equilibrium at the location in phase space where they reside. As in other particle schemes \cite{bird}, in the interest of computational efficiency, each {\it computational} deviational particle represents an {\it effective number} $N_\textrm{eff}$ of physical deviational particles. 

A dilute gas in equilibrium is described by a Maxwell-Boltzmann distribution, 
leading to a local equilibrium distribution
\begin{equation}
f^{loc}=\frac{n_{loc} }{\pi^{3/2}c_{loc}^3}\exp\left(-\frac{({\bf c}-{\bf u}_{loc})^2}{c_{loc}^2}\right)
\end{equation}
that is parametrized by the local number density $n_{loc}=n_{loc}({\bf r},t)$, the local flow velocity ${\bf u}_{loc}={\bf u}_{loc}({\bf r},t)$, and the most probable speed $c_{loc}=\sqrt{2k_bT_{loc}/m} $ based on local temperature $T_{loc}=T_{loc}({\bf r},t)$. Here, $k_b$ is Boltzmann's constant and $m$ is the molecular mass. 
In the work that follows, the underlying equilibrium distribution ($f^e$) will be identified with absolute equilibrium  
\begin{equation}
f^{e}\equiv F=\frac{n_0}{\pi^{3/2} c_0^3}\exp\left(-\frac{\mathbf{c}^2}{ c_0^2}\right)
\end{equation}
where $n_0$ is a reference (equilibrium) number density and $c_0=\sqrt{2k_bT_0/m}$ is the most probable molecular speed based on the reference temperature $T_0$. This choice provides a reasonable balance between generality, computational efficiency and simplicity. Other choices are of course possible and, depending on the problem, perhaps more efficient. However, care needs to be taken if a spatially varying or time dependent underlying equilibrium distribution is chosen since this results in a more complex algorithm \cite{thomas1,thomas2}.

Particle methods typically solve the Boltzmann equation  by applying a splitting scheme,\footnote{Note that a symmetrized algorithm provides higher-order accuracy--see Ref. \cite{pof2006} and references therein.} in which molecular motion is simulated 
as a series of collisionless advection and collision steps of length
$\Delta t$.  In such a scheme, the collisionless advection step integrates
\begin{equation}
\frac{\partial f}{\partial t} +
    \mathbf{c} \cdot \frac{\partial f}{\partial \mathbf{r}}
    = 0
\label{advection}
\end{equation} 
by simply advecting particles for a timestep $\Delta t$, 
while the collision step integrates
\begin{equation}
  \frac{\partial f}{\partial t} = \dfcoll
\label{collision}
\end{equation}
by changing the distribution by an amount $\dfcoll({\bf r},{\bf c},t)\Delta t$. Spatial discretization is introduced by treating collisions as spatially homogeneous  within (small) computational cells of volume ${\cal V}_{cell}$.  Our approach retains this basic structure; the particular form of these steps can be summarized as follows.

{\it Advection step:} It can be easily verified that when the underlying equilibrium distribution is not a function of space or time, as is the case here, the advection step for deviational particles is identical to that of physical particles [i.e. $f^d$ is also governed by equation (\ref{advection}) during the advection step]. Boundary condition implementation, however, differs somewhat because the mass flux to boundaries is now split into a deviational contribution and an equilibrium contribution. A more extensive discussion, as well as algorithmic details can be found in \cite{lowell,thomas2}.

{\it Collision step:} 
The variance-reduced form of (\ref{collision}) can be written as 
\begin{equation}
\dfcoll = \frac{1}{\tau} \left(f^{loc}-F\right)-\frac{1}{\tau} f^d
\label{start}
\end{equation}
{\it Within each computational cell} we integrate equation (\ref{start}) using a two-part process.  This integration requires local (cell) values of various quantities, denoted here by hats, which are updated every timestep by sampling the instantaneous state of the gas.

In the first part we  remove a random sample of particles by deleting particles with probability $\Delta t/\hat{\tau}$ to satisfy
\begin{equation}
\tilde{f^{d}}(t+\Delta t) = \hat{f}^{d}(t) - \frac{\Delta t}{\hat{\tau}} \hat{f}^d(t) 
\label{del}
\end{equation}
In our implementation, this is achieved through an acceptance-rejection process which can also treat the case $\hat{\tau}=\hat{\tau}({\bf c})$. 

In the second part, we 
create a set of positive and negative particles (using an acceptance-rejection process) to satisfy
\begin{equation}
\hat{f}^{d}(t+\Delta t) = \tilde{f^{d}}(t+\Delta t)  +\frac{1}{\hat{\tau}} \left[\hat{f}^{loc}(t)-F\right]\, \Delta t
\label{add}
\end{equation}
This step can be achieved by the following procedure. Let $c_{c}$ be a (positive) value such that $\hat{f}^{loc}({\bf c})-F({\bf c})$ is negligible for $||\mathbf{c}||_1 > c_{c}$, where $||\cdot||_1$ is an $L^1$-norm.  Furthermore, let $\Delta_{max}$ bound $|\hat{f}^{loc}({\bf c})-F({\bf c})|$ from above. Then, repeat  $\mathcal{N}_c $ times:
\begin{enumerate}
\item Generate uniformly distributed, random velocity vectors $\mathbf{c}$ with $||\mathbf{c}||_1<c_{c}$.
\item If $|\hat{f}^{loc}({\bf c})-F({\bf c})| > \mathcal{R} \Delta_{max}$, create a particle with velocity $\bf{c}$, at a randomly chosen position within the cell and sign $\mathrm{sgn}[\hat{f}^{loc}({\bf c})-F(\bf{c})]$.  Here, $\mathcal{R}$ is a random number uniformly distributed on [0,1]. 
\end{enumerate}
To find $\mathcal{N}_c$ we note that the number of particles (of all velocities and signs) that should be generated in a cell to obtain the proper change in the distribution function is
\begin{equation}
  \frac{\Delta t}{N_\textrm{eff}} \int_{{\cal V}_{cell}}\int_{{\bf R}^3} \frac{|f^{loc}-F|}{\tau} d^3\mathbf{c} d^3\mathbf{r}=\frac{\Delta t\mathcal{V}_{cell}}{N_\textrm{eff}\hat{\tau}} \int_{{\bf R}^3} |\hat{f}^{loc}-F | d^3\mathbf{c} 
\end{equation}
where $\mathcal{V}_{cell}$ is the cell volume.
The (expected) total number of particles ultimately generated by the above algorithm is 
\begin{equation}
\mathcal{N}_c\frac{\int_{{\bf R}^3}|\hat{f}^{loc}-F | d^3\mathbf{c}}{ \Delta_{max} 8 c_{c}^3}
\end{equation}
By equating the two expressions we obtain $\mathcal{N}_c =8 \Delta t\, \Delta_{max} \mathcal{V}_{cell} c_{c}^3/(\hat{\tau} N_\textrm{eff})$.

We have verified the above algorithm using a variety of test cases; some representative results are presented below. Figure \ref{heat} shows a comparison between numerical solution of the BGK model of the Boltzmann equation \cite{cercheat} and our simulation results for the heat flux, $q$, between two parallel, infinite, fully-accommodating plates at slightly different temperatures ($T_0$ and $T_1=T_0+\Delta T,\,\Delta T>0$) and a distance $L$ apart. The figure compares the heat flux normalized by the free-molecular (ballistic) value $|q_{fm}|=P_0c_0\Delta T/(\sqrt{\pi} T_0)$ as a function of a Knudsen number \cite{pof2006} $k=c_0\tau/L$, where $P_0=n_0k_bT_0$ is the equilibrium gas pressure.  The agreement is excellent. The simulations used approximately 50,000 particles yielding a relative statistical uncertainty\footnote{The relative statistical uncertainty  of a fluctuating hydrodynamic quantity is defined \cite{jcp} as the standard deviation of this quantity normalized by its characteristic value} of less than 0.5\%. These simulations were performed at $\Delta T=2$K, although as shown below, the cost is expected to be independent of the magnitude of $\Delta T$ in the limit of small deviation from equilibrium. We also performed DSMC simulations of the BGK model using $\Delta T=10$K and otherwise identical discretization and sampling parameters; $\Delta T=10$K was chosen as a compromise between best performance and a deviation from equilibrium that is small enough for the linearized conditions, and thus the benchmark results of \cite{cercheat}, to be valid.

\begin{figure}
 \begin{picture}(200,320)(-100,0)
\put(0,0){\includegraphics{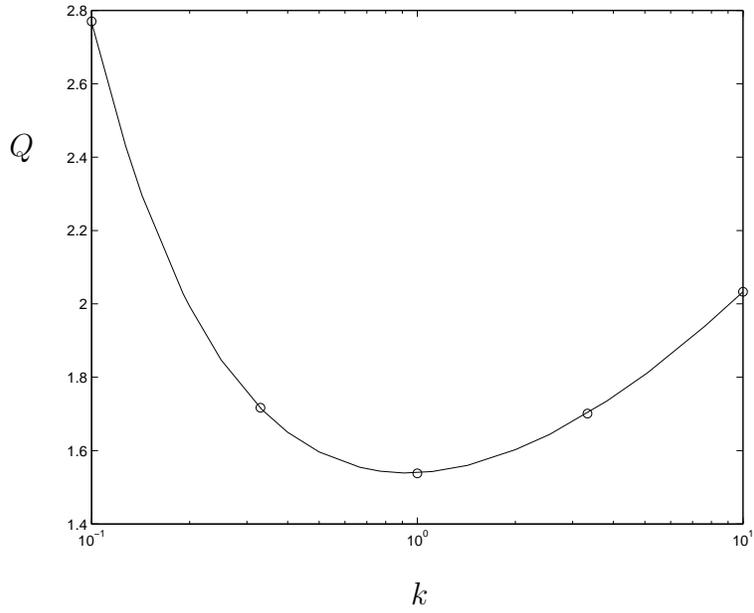}}
\put(-5,240){$Q$}
\put(147,70){$k$}
\end{picture}
  \caption{Comparison between the numerical solution of Cercignani and Daneri \cite{cercpoi} and our simulation results (circles) for the normalized flow rate in pressure-driven flow between two parallel, infinite, fully-accommodating plates. }
 \label{poi}
\end{figure}

Figure \ref{poi} shows a comparison between a numerical solution of the linearized BGK model of the Boltzmann equation \cite{cercpoi}  and our simulation results for pressure-driven flow (for small pressure gradients). Under linear conditions, pressure-driven flow can be described \cite{cercpoi}  by
\begin{equation}
\frac{\partial f^d}{\partial t} + c_x \cdot \frac{\partial f^d}{\partial x} = \dfcoll- \kappa c_z F
\end{equation}
where $\kappa=(1/P_0)\, dP/dz$ is the normalized pressure gradient (here assumed to be in the $z-$direction), and $x$ is the channel transverse direction. For $\kappa <0$, the term $-\kappa c_z F$ can be included in our formulation as a source of $|\kappa| \Delta t \int_{c_z>0} c_zF d^3{\bf c} =|\kappa| \Delta t n_0 c_0/(2\sqrt{\pi}) $ positive and an equal number of negative deviational physical\footnote{To obtain the number of computational particles we divide by $N_\textrm{eff}$.} particles per unit volume drawn from the distribution $2 \sqrt{\pi} |c_z|F/c_0$ for $c_z\geq 0 \,(-\infty<c_x,c_y<\infty)$ and $c_z< 0 \,(-\infty<c_x,c_y<\infty)$, respectively. Figure \ref{poi} shows the normalized flowrate $Q=-\rho \bar{u} c_0/(P_0\kappa L)$ as a function of $k$, where $\rho$ is the gas density and  $\bar{u}$ is the average flow velocity (averaged across the channel width). Excellent agreement is observed. 

\begin{figure}
 \begin{picture}(200,320)(-100,0)
\put(0,0){\includegraphics{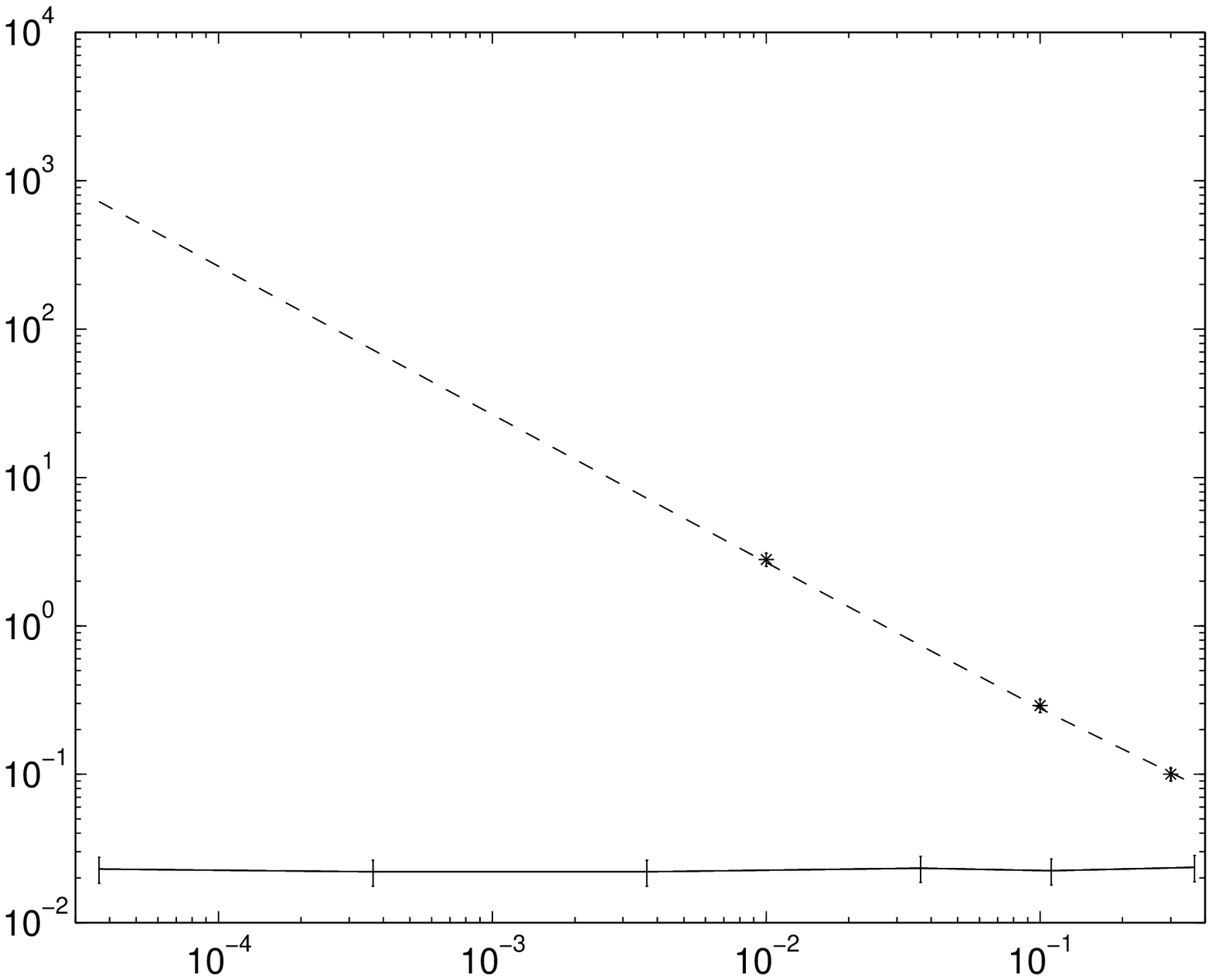}}
\put(-22,240){$\sigma_{T}$}
\put(-35,225){$\overline{T_1-T_0}$}
\put(120,220){DSMC}
%\put(130,70){$(T_1-T_0)/T_0$}
\put(132,70){$\underline{T_1-T_0}$}
\put(145,55){$T_0$}
\end{picture}
  \caption{The relative statistical uncertainty in temperature  as a function of $(T_1-T_0)/T_0$ for $T_0=273$K. Simulation results  are presented  and compared to DSMC (indicated) which serves as a canonical case of a ``non-deviational'' method. The dashed line is obtained from the theory presented in \cite{jcp}, while stars denote actual DSMC simulation results}
 \label{fluct}
\end{figure}

As stated above and shown in \cite{pof2005,lowell,thomas2}, this class of deviational methods exhibit statistical uncertainties that scale with the local deviation from equilibrium thus allowing the simulation of arbitrarily low deviations from equilibrium at a cost that is independent of this deviation. Here we demonstrate this feature by studying the statistical uncertainty of the temperature in a problem involving heat transfer. Specifically, figure \ref{fluct} shows the relative statistical uncertainty in the temperature ($\sigma_T$) as a function of the normalized wall temperature difference $(T_1-T_0)/T_0$ in the heat transfer problem discussed above, for $k=1$ and $T_0=273$K; in evaluating $\sigma_T$, the characteristic value for temperature was taken to be the difference $T_1-T_0$.  The standard deviation is measured from two computational cells in the middle of the computational domain, each  containing approximately 950 particles. The figure shows that, for small $T_1-T_0$, the relative statistical uncertainty remains independent of this quantity in sharp contrast to ``non-deviational'' methods\footnote{The statistical uncertainty of ``non-deviational'' methods (for the same number of particles per cell) is estimated using equilibrium statistical mechanics \cite{jcp}. This approximation is very reasonable for small deviations from equilibrium \cite{jcp}.}. Moreover, the variance reduction achieved is such that significant computational savings are expected for $(T_1-T_0)/T_0\lesssim 0.1$. 

The algorithm described above imposes no restrictions on the magnitude of $f^d$, although it is expected that the deviational approach will be significantly more efficient than traditional approaches when $f^d$ is small. If $f^d$ is sufficiently small for linearization to be appropriate, under some conditions, significant gains in computational efficiency can be achieved by taking the following into consideration. Under linear conditions, for the present model, we can write
\begin{equation}
f^{loc}-F=F\left[\omega+2\frac {{\bf c}\cdot {\bf u}_{loc}}{c_0^2}+\left(\frac{{\bf {c}}^2}{c_0^2}-\frac{3}{2}\right)\phi\right]
\label{lin}
\end{equation}
where $\omega=n_{loc}/n_0-1$ and $\phi=T_{loc}/T_0-1$. This representation can be very useful for improving the computational efficiency of update (\ref{add}). For example, for isothermal constant density flows, particles can be generated from a combination of a normal distribution and analytic inversion of the cumulative distribution function, which is significantly more efficient than acceptance-rejection. Alternatively, (\ref{lin}) provides a means of obtaining tight bounds for $|f^{loc}-F|$ and thus reducing the number of rejections if the acceptance-rejection route is followed.

We have presented an efficient variance-reduced particle method for solving the Boltzmann equation in the relaxation-time approximation.  The method combines simplicity with a number of desirable properties associated with particle methods, such as robust capture of traveling discontinuities in the distribution function and efficient collision operator evaluation using importance sampling \cite{pof2005}, without the high relative statistical uncertainty associated with traditional particle methods in low-signal problems. In particular, the method presented here can capture arbitrarily small deviations from equilibrium for constant computational cost. More sophisticated techniques with spatially variable underlying equilibrium distribution \cite{thomas1,thomas2} are expected to increase computational efficiency by reducing the number of deviational particles required to describe the local state of the gas. One such technique is described in \cite{pre}. 
\section*{Acknowledgements}
This work was supported by the Singapore-MIT Alliance under the HPCES program.

\newpage

\begin{figure}
 \begin{picture}(200,320)(-100,0)
\put(0,0){\includegraphics{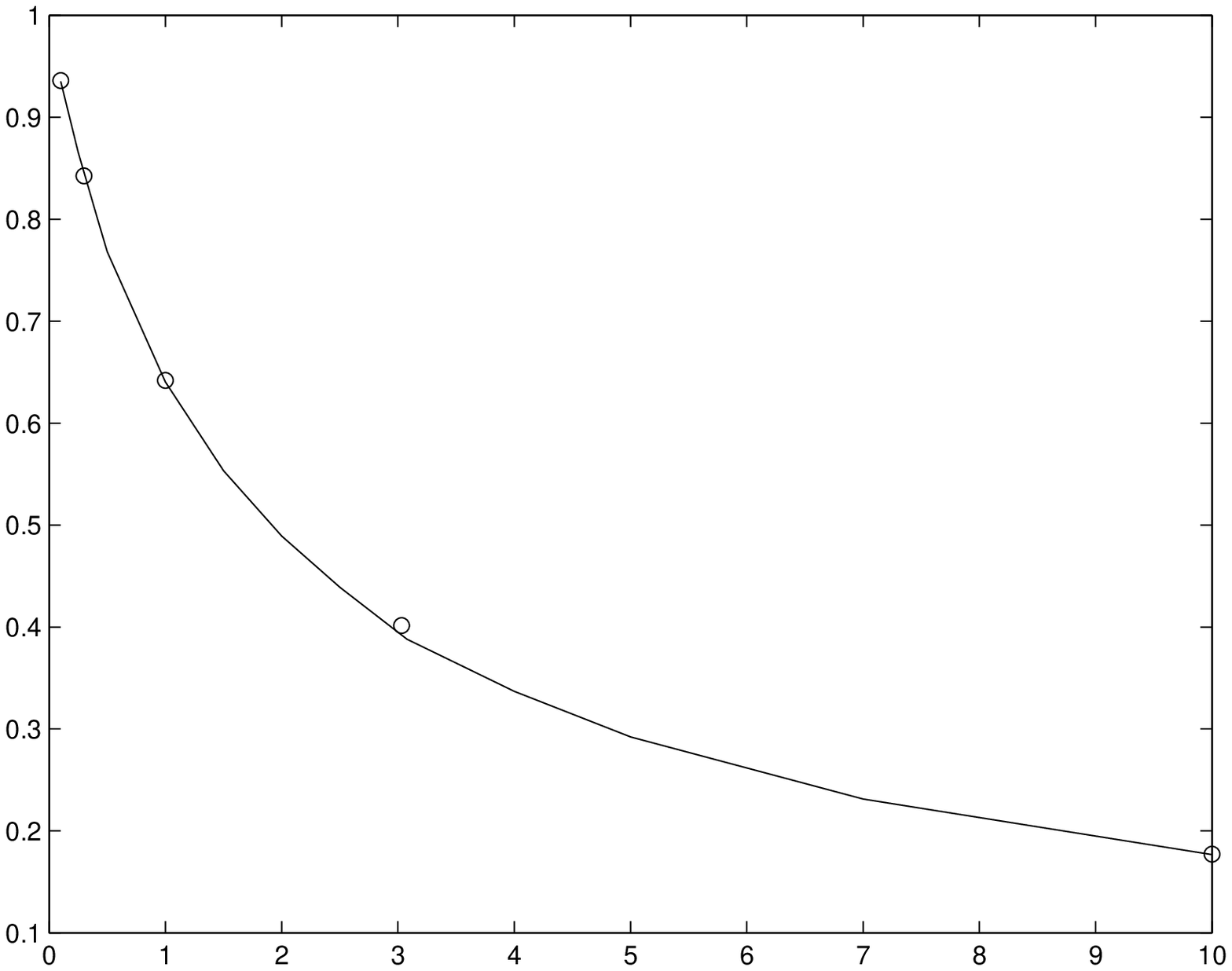}}
\put(-25,240){$q/q_{fm}$}
\put(140,70){$1/k$}
\end{picture}
  \caption{Comparison between the numerical solution of Bassanini et al. \cite{cercheat} and our simulation results (circles) for the heat flux between two parallel, infinite, fully-accommodating plates. Some numerical solution data for $k>1$ have been transcribed from Ref. \cite{cerc}.}
 \label{heat}
\end{figure}
\newpage

  \newpage

\end{document}